\documentclass[s]{iucr}                        
\newcommand{\text}{\rm}

\usepackage{graphicx}

\journalcode{S}              

\begin{document}                  


\title{The 1-Megapixel pnCCD Detector for the Small Quantum Systems Instrument at the 
		European XFEL: System and Operation Aspects}
\shorttitle{XFEL 1-Mpix pnCCD System and Operational Aspects}


\cauthor[a]{Markus}{Kuster}{markus.kuster@xfel.eu}{}
\author[a]{Karim}{Ahmed}
\author[a]{Kai-Erik}{Ballak}
\author[a]{Cyril}{Danilevski}\aufn{On leave from the European XFEL, Schenefeld, Germany}
\author[a]{Marko}{Ekmed\v{z}i\'{c}}
\author[a]{Bruno}{Fernandes}
\author[a]{Patrick}{Gessler}
\author[b]{Robert}{Hartmann}
\author[a]{Steffen}{Hauf}
\author[b]{Peter}{Holl}
\author[a]{Michael}{Meyer}
\author[a]{Jacobo}{Monta\~no}
\author[a]{Astrid}{M\"unnich}
\author[a]{Yevheniy}{Ovcharenko}
\author[a]{Nils}{Rennhack}
\author[a]{Tonn}{R\"uter}
\author[c,d]{Daniela}{Rupp}
\author[b]{Dieter}{Schlosser}
\author[a]{Kiana}{Setoodehnia}
\author[a]{R\"udiger}{Schmitt}
\author[b]{Lothar}{Str\"uder}
\author[c,e]{Rico Mayro P.}{Tanyag}
\author[e]{Anatoli}{Ulmer}
\author[a]{Hazem}{Yousef}
  
\aff[a]{European XFEL, Holzkoppel 4, \city{22869 Schenefeld} \country{Germany}}
\aff[b]{PNSensor GmbH, Otto-Hahn-Ring 6, \city{81739 M\"unchen} \country{Germany}}
\aff[c]{Max-Born-Institute, Max-Born-Straße 2A, \city{12489 Berlin} \country{Germany}}
\aff[d]{LFKP. ETH Z\"urich, John-von-Neumann-Weg 9, \city{8093 Z\"urich} \country{Switzerland}}
\aff[e]{IOAP, Technische Universit\"at Berlin, Hardenbergstraße 36, \city{10623 Berlin}\country{Germany}}






\keyword{X-ray Detector} 
\keyword{Photon Detection} 
\keyword{pnCCD}
\keyword{X-ray CCD} 
\keyword{Free Electron Laser}



\maketitle                        

\begin{synopsis}
	A description of the 1-Megapixel pnCCD detector design and capabilities, 
	its implementation at EuXFEL's Small Quantum Systems Instrument both mechanically 
	and from the controls side as well as important data correction steps aim to provide 
	useful background for users planning and analyzing experiments at EuXFEL and may serve 
	as a benchmark for comparing and planning future endstations at other FELs.
\end{synopsis}

\begin{abstract}
	The X-ray free-electron lasers that became available during the last decade, 
	like the European XFEL (EuXFEL), place high demands on their instrumentation. 
	Especially at low photon energies below $1\,\text{keV}$, detectors with 
	high sensitivity, and consequently low noise and high quantum efficiency, are 
	required to enable facility users to fully exploit the scientific potential of 
	the photon source. 
	A 1-Megapixel pnCCD detector with a $1024\times 1024$ pixel format has been 
	installed and commissioned for imaging applications at the Nano-Sized Quantum 
	System (NQS) station of the Small Quantum System (SQS) instrument at EuXFEL. 
	The instrument is currently operating in the energy range between $0.5$ and 
	$3\,\text{keV}$ and the NQS station is designed for investigations of the 
	interaction of intense FEL pulses with clusters, nano-particles and small 
	bio-molecules, by combining photo-ion and photo-electron spectroscopy with 
	coherent diffraction imaging techniques.
	The core of the imaging detector is a pn-type charge coupled device (pnCCD) 
	with a pixel pitch of $75\,\mu\text{m}\times 75\,\mu\text{m}$. Depending on the 
	experimental scenario, the pnCCD enables imaging of single photons thanks to its 
	very low electronic noise of $3\,\text{e^-}$ and high quantum efficiency. Here 
	we present an overview on the EuXFEL pnCCD detector and the results from the 
	commissioning and first user operation at the SQS experiment in June 2019. The 
	detailed descriptions of the detector design and capabilities, its 
	implementation at EuXFEL both mechanically and from the controls side as well 
	as important data correction steps aim to provide useful background for users 
	planning and analyzing experiments at EuXFEL and may serve as a benchmark for 
	comparing and planning future endstations at other FELs.	 
\end{abstract}


\section{Introduction}

The increasing availability of X-ray free-electron laser (FEL) sources 
\cite{Altarelli:2006a,Emma:2010a,Ganter:2010a,Ishikawa:2012a,Kang:2017a} over 
the last decade has triggered various new developments and improvements in 
spectrometer and detector technologies in order to enable the scientific 
community to fully exploit the capabilities of these sources for experimental 
investigations. 

In the area of detector technology, silicon based direct detection 
hybrid pixel detectors (HPD) with up to $4.5$ MHz imaging capability became 
available during the last years, mainly enabled by the rapid advancement in 
microelectronics technology. A few examples are integrating detectors like the 
Large Pixel Detector (LPD) \cite{Hart:2012}, the Adaptive
Gain Integrating Pixel Detector (AGIPD) \cite{Allahgholi:2019a} and the DEPFET 
Sensor with Signal Compression (DSSC) \cite{Porro:2012a}. Being the result of a 
more than ten years long development effort, these technologies have the potential 
to open new avenues for scientific applications at FELs.

Especially scientific applications relying on very low photon detection noise 
and/or spectro-imaging capability benefit from the continuous improvements of 
the Charge Coupled Device (CCD) technology during the last decade. The FastCCD 
\cite{Doering:2012a}, MPCCD \cite{Kameshima:2014a} and pnCCD \cite{Strueder:2010a}
are a few exemplary monolithic detectors playing an important role at storage
rings and FELs since years (see \citeasnoun{Graafsma:2020a} and \citeasnoun{Hatsui:2015a}
for an in-depth review).

The European XFEL started user experiments in 2017 \cite{Decking:2020a} and is 
operating in the soft X-ray and hard X-ray regime, i.e. at photon energies from 
$270\,\text{eV}$ up to more than $20\,\text{keV}$. Presently six scientific 
instruments \cite{Tschentscher:2017a} are providing high-performance 
experimental tools for a broad range of scientific applications in research 
fields like fundamental research on atoms and molecules, chemistry, solid state 
physics and biology.

One of these instruments, the Small Quantum Systems (SQS) instrument, is 
dedicated to investigations of non-linear and time-resolved phenomena on atomic 
and molecular systems, as well as to studies of clusters and nano-objects under 
irradiation with ultra-short, highly intense soft X-ray pulses. The SQS 
scientific instrument is located behind the SASE~3 soft X-ray undulator, which 
delivers intense, ultra-short (about $25\,\text{fs}$) and spatially coherent 
pulses of X-ray light, currently with photon energies in the range from $0.5$ to 
$3.0\,\text{keV}$ with smooth wavelength tunability and pulse energies up to 
$5\,\text{mJ}$.

Three versatile and interchangeable experimental stations are available at the 
SQS instruments for the user community. One of them, the Nano-sized Quantum 
Systems (NQS) station (see Fig.~\ref{Fig:pnCCD-NQS-Chamber}), is designed for 
investigations of the interaction of intense FEL pulses with clusters, 
nano-particles and small bio-molecules by combining photo-ion and photo-electron 
spectroscopy with coherent diffraction scattering techniques. Experiments based 
on electron and ion spectroscopy are realized by using velocity-map imaging 
(VMI) \cite{Eppink:1997a} and time-of-flight (TOF) mass spectrometers 
\cite{Wiley:1955a}, respectively. For experiments on coherent diffraction 
imaging a 1-Megapixel pnCCD detector is available since June 2019.

\section{The pnCCD Camera System}

The pnCCD technology was originally developed for spectro-imaging applications 
in X-ray Astronomy \cite{strueder:90a,strueder:01a} and is used in this field 
for more than 20~years \cite{xmm:2020a}. Since then, \mbox{pnCCDs} have been 
successfully deployed for a broad range of scientific applications with 
significantly different scientific requirements. This includes imaging of 
electrons \cite{Ryll:2016a}, as well as of photons from visible light 
to X-rays \cite{Strueder:2010a}, soft gamma ray energies and indirect detection 
of dark matter \cite{kuster:07c}. Depending on the experimental scenario, imaging 
of single photons is combined with high-resolution spectroscopy.


The design of the 1-Megapixel pnCCD camera system for the SQS instrument
is optimized to fulfill the technical and scientific requirements of 
the NQS endstation, while maintaining the flexibility to move the camera as a 
self contained unit between different experiments at the European XFEL. The core 
component of the detector is a fully depleted pnCCD. A valuable feature of 
this kind of sensor is its integrated front-end electronics and fully column 
parallel readout providing excellent imaging and spectroscopic performance, 
long term stability, and high speed readout.

\subsection{Detector Mechanics, Vacuum and Cooling System}

\begin{figure}
  \includegraphics[width=0.49\columnwidth]{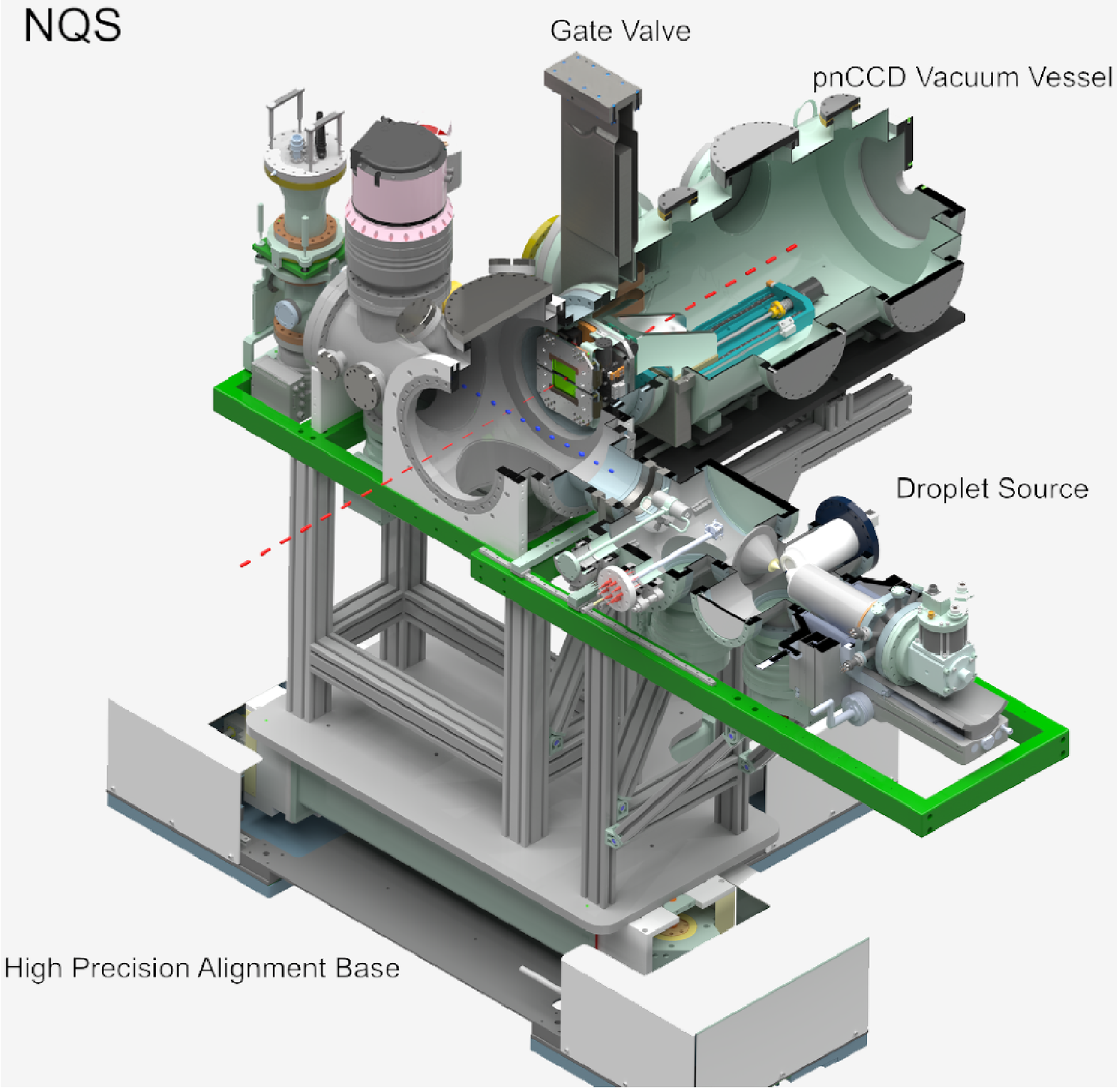}
  \hfill
  \includegraphics[width=0.49\columnwidth]{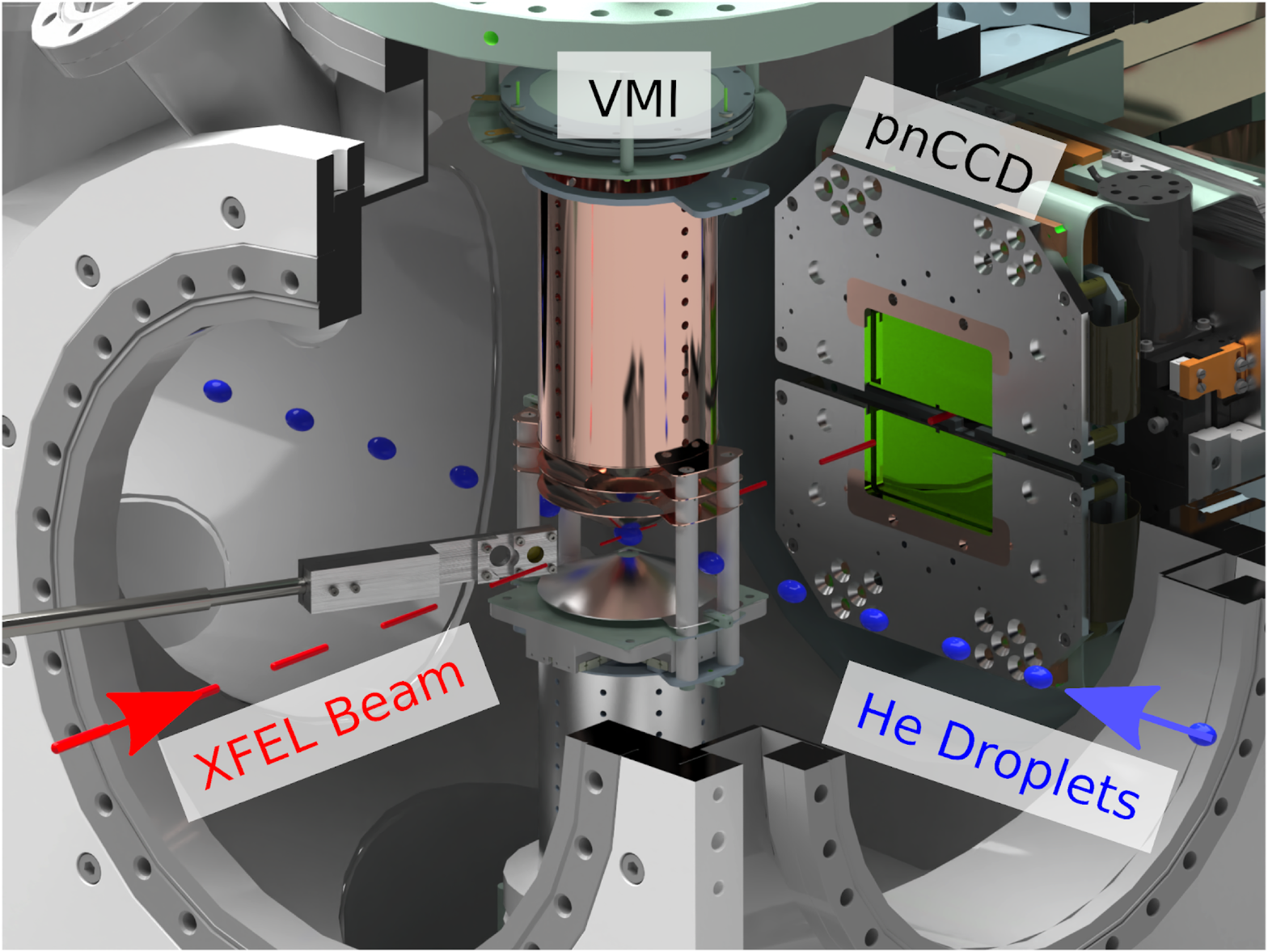}
  \label{Fig:pnCCD-NQS-Chamber}
  \caption{Left: Overview of the NQS experimental chamber at the SQS 
	instrument. Right: Enlarged view of the interaction region showing the pnCCD
	detector mounted in the direction of the photon beam. The photon beam enters 
	the NQS chamber from the bottom left (red dashed line) and the droplet source injects 
	the sample from the bottom right (blue droplets).}
\end{figure}

Figure~\ref{Fig:pnCCD-NQS-Chamber} shows a schematic cross section of the pnCCD 
detector system with its subcomponents as implemented at the NQS endstation. The
pnCCD detector is mounted in beam direction at a distance of about $30\,\text{cm}$
behind the interaction volume defined by the crossing of the FEL (dashed red line) and the sample 
delivery axis (blue dotted line) (Fig.~\ref{Fig:pnCCD-NQS-Chamber}). The FEL beam is focused down to a spot 
size of about $1.5\times 1.5\,\mu\text{m}^2$ (FWHM) by a pair of highly 
polished Kirkpatrick-Baez (KB) mirrors. The NQS chamber is kept at a vacuum level
of $10^{-8}$ to $10^{-9}\,\text{mbar}$ in order to operate the electron and ion 
spectrometer. A set of apertures is installed at the entrance of the chamber allowing
to define the FEL beam profile and to reduce stray light on the photon detector.

A DN~300 shutter connecting the pnCCD camera's vacuum chamber to the NQS endstation 
can be closed to seal it off hermetically, if either the detector system is operated
stand-alone (e.g. for system qualification and calibration) or alignment procedures of 
the target beam are carried out that may bear a risk for the pnCCD detector. The design 
of the ultra high vacuum system of the camera and detector components enables operation 
in the low $10^{-8}\,\text{mbar}$ pressure range. 

The pnCCD sensor plane and its cooling infrastructure are mounted on a  
$300\,\text{mm}$ long and UHV compatible linear stage, enabling the flexibility of  
changing the distance between the sample and the sensor plane along the FEL beam 
axis (i.e. in {$z$-direction} when referring to beam line coordinates, see 
Fig.~\ref{Fig:pnCCD-NQS-Chamber} and \citeasnoun{Sinn:2013a}). In the fully
retracted position, the sensor plane is at a distance of $350\,\text{mm}$
to the beam focus. At the closest distance to the beam focus, i.e. at a distance
of $50\,\text{mm}$, the sensor plane sticks out of detector's vacuum chamber into
the NQS chamber. Positioned to those most extreme positions, the sensor plane covers 
scattering angles of up to $9.5^\circ$ and $49.4^\circ$, respectively. 

\begin{figure}
	\centering
 	\includegraphics[width=0.97\columnwidth]{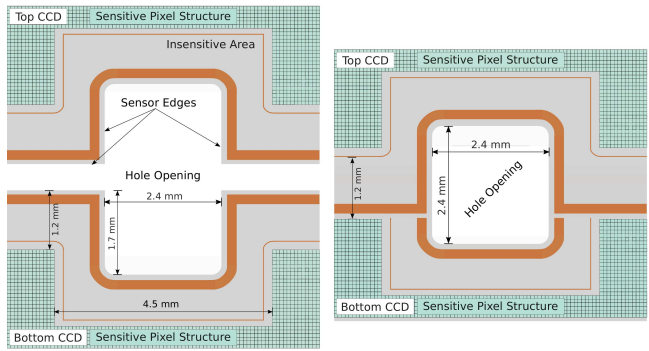}
 	\label{Fig:Sensor-Hole-Geometry}
 	\caption{Schematic view of the sensor highlighting the region close to the central 
 		rectangular hole. The sensor and hole geometry are shown for two configurations.  
 		Left: The top and bottom CCD are moved up and down into an open hole configuration. 	
 		Right: The smallest hole configuration is shown. Due to the staggered arrangement of 
 		the upper and lower CCD, the upper CCD is overlapping the lower 
 		CCD such that the slit is completely closed and a $2.4 \,\text{mm}\times 2.4\,\text{mm}$ 
 		large quadratic hole remains in the center of the sensor plane.}
\end{figure}

Two vacuum motion stages allow separation of the sensor halves in vertical 
($y$) direction with a precision of $1\,\mu\text{m}$, i.e. perpendicular to 
the FEL beam. The resulting gap between the two halves of the pnCCD sensor 
is up to $40\,\text{mm}$ wide, allowing the user to enlarge the detection plane 
in order to cover a larger $q$ range. Passage of the primary FEL beam is enabled by a laser cut, 
$2.4\,\text{mm}\times 2.4\,\text{mm}$ large, rectangular hole as illustrated 
in Fig.~\ref{Fig:Sensor-Hole-Geometry}. The slit and the hole can be completely 
closed, as the two sensor planes can be staggered leaving a gap of $3\,\text{mm}$ 
between the sensor planes in beam direction, with the top half covering the
lower half. When the slit is completely closed, a $1.2\,\text{mm}$ insensitive area
remains between the lowest pixel column of the top detector half and the uppermost 
pixel column of the bottom half, whereby no part of the sensitive sensor 
area is blocked. 

The significantly reduced thermally generated dark-current of the latest 
generation pnCCDs in comparison to earlier versions \cite{Holl:2006a}, allows 
operating the pnCCD sensors at temperatures up to $-20^\circ\,\text{C}$ while 
achieving an electronic noise level of $3\,\text{e}^-$ rms in high gain and 
about $10\,\text{e}^-$ rms in low gain. To achieve the best imaging and 
spectroscopic performance, temperature stability of the sensor better than $\pm 
1\,\text{K}$ is required.

\begin{figure}
	\centering
  	\includegraphics[width=0.77\columnwidth]{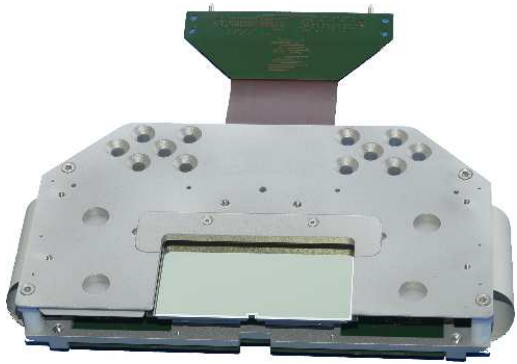}
  	\label{Fig:Sensor-Hybrid-Board}
	  	\caption{Image of the pnCCD sensor hybrid board. Two of these boards are installed in the 
	  	1 Mpix camera. Each board carries one sensor half, $512\times 1024\,\text{pixels}$ large
	  	and the in vacuum readout electronics. The flex lead shown on the top of the image, 
	  	provides the electronic interface to the part of the readout electronics installed outside 
	  	the vacuum chamber.}
\end{figure}

The thermal power budget of the sensor plane is dominated by the thermal 
dissipation power of the $16$ Application Specific Integrated Circuit (ASIC) 
chips mounted next to the pnCCD sensors on the sensor hybrid board (see 
Section~\ref{Sec:Data Readout and Processing}, Fig.~\ref{Fig:Sensor-Hybrid-Board} 
and Fig.~\ref{Fig:Sensor-Geometry}). During readout of the pnCCD, each of the 
ASICs contributes with $0.3\,\text{W}$ to the total power budget of 
$5\,\text{W}$. In order to reduce the power dissipation to a minimum, 
the ASICs are switched to stand-by when not needed for signal processing, 
e.g. during the integration phase of the CCD's image taking cycle.

The detector's active cooling system is optimized to the above mentioned 
requirements. It is seamlessly integrated into the camera and the motion system. 
Rigid high-conductivity copper connections transfer the heat from the Al-Si alloy 
sensor frame to the rear side of the sensor plane. Subsequently, four 
flexible copper braids transmit the dissipated thermal power to the in-vacuum 
cold fingers of two Polycold\textsuperscript{\textregistered} Compact Coolers 
(PCC). In the configuration used for this application, the PCCs provide a 
maximum cooling power of $\approx 24\,\text{W}$. 
With this design, the sensor temperature is stable to better than $\pm 
0.1\,\text{K}$. 
This temperature tolerance is sufficient in significantly minimizing the detector's 
dark current and noise, as needed in photon-sensitive X-ray imaging. The time required to cool the pnCCDs in vacuum from room temperature to the nominal operating 
temperature of $-30^\circ\,\text{C}$, and to reach stable thermal operating conditions 
is about $3\,\text{h}$. Temperature stability over a time scale of days within 
the above mentioned requirement has been demonstrated with this design.

During detector operation, the temperature of the pnCCD sensor is 
continuously monitored via four platinum resistance thermometers (type PT~1000) mounted on 
the cold fingers and the copper block attached to the pnCCD modules. The temperature
readings, amongst other slow control values,  are logged by the Karabo control system 
of the European XFEL \cite{hauf:2019a}. The pnCCD calibration pipeline can use these 
temperature values as reference for data correction and processing (see Sec.~\ref{Sec:Data Readout and Processing}).

\subsection{pnCCD Sensor and Signal Processing}
\label{Sec:pnCCD Sensor and Signal Processing}
\begin{figure}
	\centering
  	\includegraphics[width=.97\columnwidth]{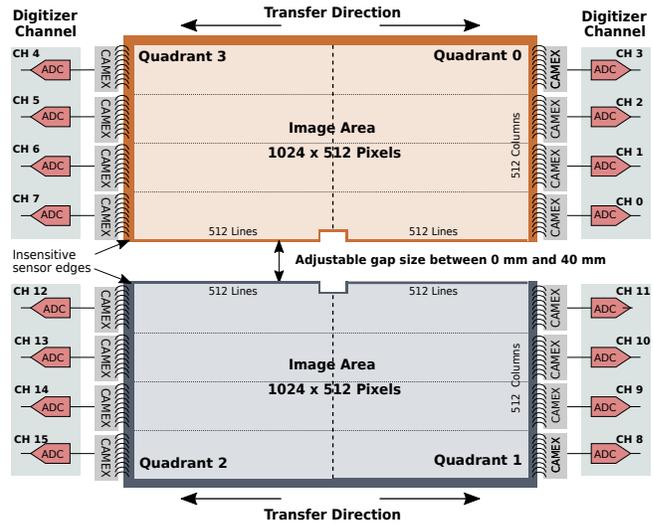}
  	\label{Fig:Sensor-Geometry}
  	\caption{Schematic view of the pnCCD sensor geometry as seen when looking in 
  	beam direction onto the detector (in beam line coordinates this corresponds 
  	to the positive $z$ direction), i.e. down-stream FEL beam direction. The sensor
  	is divided into two halves, which can be operated independently. The designation 
  	of the digital signal channels and detector quadrants is the same as the one 
  	used in the meta data of the Karabo online data stream and the HDF5 raw/calibrated
  	data files. Both data sources are accessible to beamline users for data analysis.}
\end{figure}

The pnCCD sensor plane has a sensitive area of 
$7.68\,\text{cm} \times 7.68\,\text{cm}$ and consists of two monolithic chips 
(sensor halves) as schematically shown in Fig.~\ref{Fig:Sensor-Geometry}. Both are identically 
designed, back-illuminated pnCCDs. Each has $1024\times 512$ pixels and a fully 
depleted, $450\,\mu\text{m}$ thick sensitive volume. The pixel size of 
$75\,\mu\text{m} \times 75\,\mu\text{m}$ allows for sub-pixel position 
resolution of better than $10\,\mu\text{m}$ (at normal incidence) in the single photon detection 
regime when charge centroiding techniques are applied~\cite{Ihle:2017a}. The two halves
are operated and read out independently in a split-frame mode 
(see Fig.~\ref{Fig:Sensor-Geometry}). Further performance parameters of the sensor and 
readout electronics are summarized in Table \ref{Table:pnCCD-Detector-Parameters}.

During the readout process, the signal charges generated by X-ray photons are 
transferred line by line along the pnCCD transfer columns. Each column is 
terminated by a charge collecting anode connected to an integrated n-channel 
junction gate field-effect transistor (JFET). This on-chip amplifier is operated 
in source follower mode and bonded to 
one input channel of the CMOS Amplifier and MultiplEX (CAMEX) ASIC 
~\cite{Buttler:88a,Herrmann:2008a}. This column-parallel signal 
amplification concept allows for high frame rates in combination with low noise performance.

In total $8$ CAMEX ASICs with $128$ inputs each are required for handling the 
signal of the $1024$ JFET source follower outputs of each pnCCD chip. For 
further signal processing, the CAMEX provides two-stage amplification, filtering, 
noise bandwidth limitation and 8-fold Correlated Double Sampling (CDS). The 
multiple baseline and signal sampling enabled by the CDS increases the signal to 
noise ratio proportional to $\sqrt{n}$, where $n$ is the number of sampling 
points. Through a digital control register 
it is possible to adjust the gain of the 
pre-amplifiers to the experimental conditions. Apart from the highest gain $g=1$, 
gain values of  $g=1/4, 1/16, 1/64, 1/256$ and $1/1024$ can be configured 
before data taking. Finally the signal is stored in a sample and hold (S\&H) 
stage before all $128$ parallel channels are multiplexed in 
$1$ analog output channel per CAMEX.

After serialization by the CAMEX, the analog signal is transferred to and 
digitized by two SIS8300 $\mu\text{TCA}$ digitizer boards providing 10~ADC 
channels with 16~bit resolution each. The EuXFEL uses these boards as a facility-wide 
standard for high speed digitizer applications. The boards sample the signal with a 
frequency between $10~\text{MHz}$ and $125~\text{MHz}$ per 
channel. Application specific readout, data formatting, data reduction and 
processing algorithms can be implemented on a Virtex~V Field Programmable Gate 
Array (FPGA) available on the SIS8300 board~\cite{Struck:2020a}. Each board 
in addition reads the pixel clock signal provided by the timing sequencer of the 
pnCCD for synchronization. The digitizer FPGA firmware is configured to sample 800~kSamples per 
CAMEX and image, as determined by the 10~Hz train trigger, which corresponds to 
1 or 2~samples per pixel and readout cycle. The CAMEX is typically operated at 
$10\,\text{MHz}$, enabling image rates up to $100$ frames per second.

Figure~\ref{Fig:Sensor-Geometry} shows the geometric arrangement of the sensor 
plane, the signal transfer direction, and illustrates how the ASIC chips are 
arranged on the right and left side of the sensor chips. The designation of  
the output channels of the ADCs and detector quadrants is the same as the one 
used in the meta data of the Karabo online data stream and the HDF5 \cite{hdf5:2020a}
raw/calibrated data files.

Visible and infrared background light, e.g. from a 
$800\,\text{nm}$ pump-laser, can significantly deteriorate the signal to noise 
ratio of an experiment, especially when aiming at photon energies $< 
1\,\text{keV}$ and small scattering signals. A blocking filter for visible and 
IR light is therefore crucial for maintaining a high signal to noise ratio in 
the presence of background signal sources. In general two technological 
solutions exist to mitigate this problem, either by using a thin and coated foil, 
mounted in front of the sensor plane, or by a filter deposited on the entrance 
window of the sensor. Appropriate foils, having a thickness of a few 
hundred nanometers, are very fragile and require a support structure, that may 
additionally scatter photons as undesired background signal.

\begin{figure}
  \includegraphics[width=0.97\columnwidth]{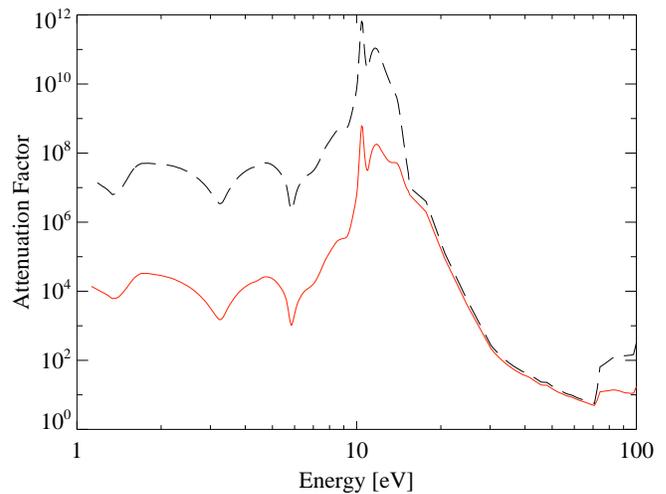}
  \label{Fig:pnCCD-UV-Optical-Filter}
  \caption{The attenuation factor of a $150\,\text{nm}$ (black dashed line) and 
  	$60\,\text{nm}$ (red line) thick aluminium, 
	$\text{SiO}_2$ and $\text{Si}_3\text{N}_4$ light blocking filter deposited on 
	the entrance window of the pnCCD sensor. The filter efficiently suppresses near 
	infrared, visible and ultaviolet radiation up to $10\,\text{eV}$ by 
	a factor $10^7$ or $10^4$, respectively. The influence of the aluminum 
	absorption decreases due to the plasma frequency cutoff at $\omega_p\approx 16\,\text{eV}$ 
	resulting in the refractive index $n$ becoming increasingly smaller between $15\,\text{eV}$ and 
	$73\,\text{eV}$, the Al L edge. 
	}
\end{figure}

\begin{figure}
  \includegraphics[width=0.97\columnwidth]{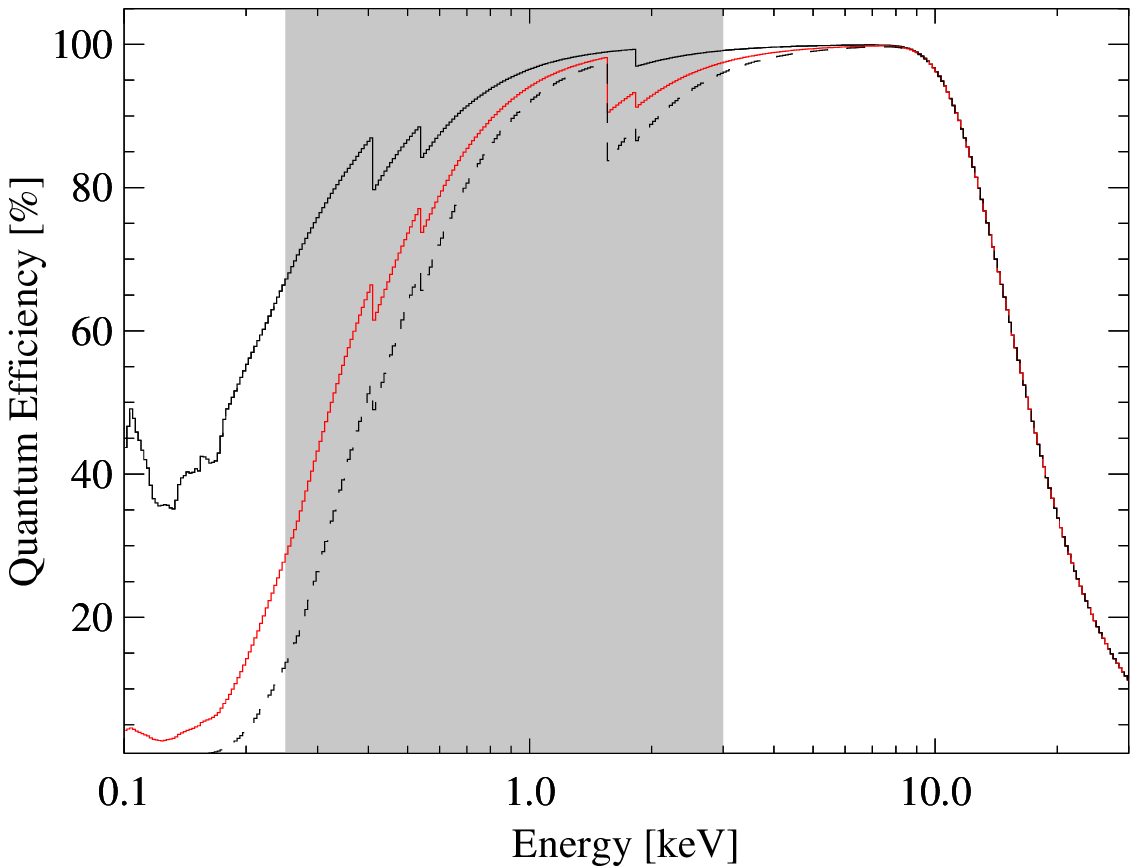}
  \label{Fig:pnCCD-QuantumEfficiency}
  \caption{The quantum efficiency of the pnCCD sensor in the energy range 
	between $100\,\text{eV}$ and $30\,\text{keV}$ is shown. The  black dashed line and the red solid line
	correspond to the quantum efficiency of a pnCCD sensor with a $150\,\text{nm}$ and  $60\,\text{nm}$ thick 	
	aluminum  entrance window, respectively. For comparison, the quantum efficiency of an entrance window without an 
	aluminum  layer is shown (solid black line). Note that in the current installation, an Al thickness of $150\,\text{nm}$ 
	is applied, while a future upgrade is planned with a $60\,\text{nm}$ thick Al coating. The grey area marks the baseline 
	photon energy available at the SQS instrument.}
\end{figure}

With a light blocking filter directly deposited on the entrance window of the 
sensor, we can circumvent these disadvantages. The design of such a filter 
requires a careful trade-off between attenuation of visible light and the 
desired X-ray quantum efficiency. For this detector, we have chosen an in total 
$150\,\text{nm}$ thin blocking filter, consisting of layers of aluminium, $\text{SiO}_2$ and 
$\text{Si}_3\text{N}_4$ directly deposited on the $\text{p}+$-doped silicon 
sensor backside. During a major upgrade planned in the second half of 2020, we will 
implement a new sensor module with an improved sensitivity for low energy photons, 
enabled by the implementation of a thinner aluminum entrance window with a thickness 
of 60 nm. The combination of the three layers efficiently attenuates 
photons with energies between $1\,\text{eV}$ and $10\,\text{eV}$ by at 
least a factor of $\approx 10^4$ using a $60\,\text{nm}$ and $\approx 10^7$ with 
a $150\,\text{nm}$ thick aluminium layer as shown in 
Fig.~\ref{Fig:pnCCD-UV-Optical-Filter}. A further benefit
of this kind of ultra-thin entrance window is its homogeneous response to low
energy photons, resulting in good energy 
resolution and high quantum efficiency of at least $75\,\%$ in the energy range 
between $0.6\,\text{keV}$ and $14\,\text{keV}$ (see Fig.~\ref{Fig:pnCCD-QuantumEfficiency}).

\subsection{pnCCD Read Out Timing and Synchronization to the XFEL Pulse Pattern}
\label{Sec:pnCCD Read Out and Synchronization to the XFEL Pulse Pattern}

\begin{figure}
  \includegraphics[width=0.97\textwidth]{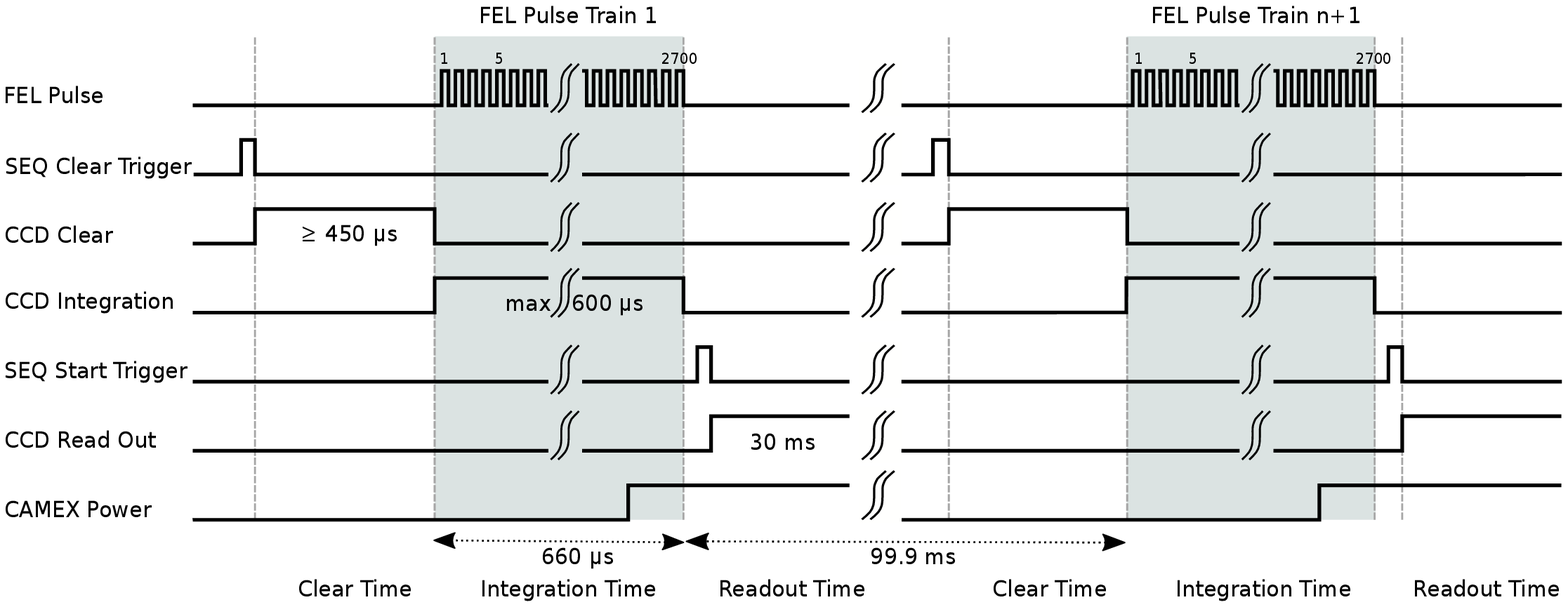}
  \label{Fig:Timing-Diagram}
  \caption{Schematic diagram illustrating the pnCCD readout timing and 
	synchronization to the EuXFEL pulse pattern. For a more detailed 
	description, we refer the reader to the text.}
\end{figure}

The EuXFEL machine delivers up to $27000$ spatially coherent X-ray pulses per second 
in a unique time structure. Ten times per second, a train of equidistantly separated 
X-ray pulses arrives at the sample interaction point in the NQS sample chamber as 
shown in the topmost line of Fig.~\ref{Fig:Timing-Diagram}. Each pulse train can contain 
between $1$ and up to $2700$ X-ray pulses with a minimum inter pulse 
separation of $220\,\text{ns}$. This corresponds to a maximum pulse repetition 
frequency of $4.5\,\text{MHz}$.

The synchronization of the detector is accomplished through the EuXFEL timing 
system~\cite{Gessler:2020a}, which provides triggers to the sequencer of the 
pnCCD. Since the sequencer uses an internal clock, the SIS8300 ADC is 
synchronized to the sequencer to ensure phase aligned readout of the detector 
signals. In this way, the detector front-end is correctly aligned with 
the timing of the machine.


Since the pnCCD is continuously biased, the sensor bulk is sensitive to visible, 
IR and X-ray photons at any time of the integration and readout cycle. As a 
consequence photons being absorbed in the depleted sensor bulk between two 
consecutive readout cycles would create a  potential background signal which 
does not originate in an interaction of the FEL beam with the sample. Therefore, 
the pnCCD sensor is cleared of charge before the arrival of the first X-ray FEL pulses by 
quickly shifting residual charges to the readout anode and discarding the charge 
signal.

This $420\,\mu\text{s}$ long clear process is followed by the signal integration 
period (see Fig.~\ref{Fig:Timing-Diagram}) with a duration chosen according to 
scientific requirements. Its minimum duration
is $600\,\text{ns}$, with an upper limit given, e.g. by the desired time 
resolution. A typical value is $600\,\mu\text{s}$ which is sufficient to record all signals produced 
by the X-ray pulses within one pulse train. The final signal readout, 
amplification, filtering and digitization take additional $14\,\text{ms}$. The 
readout sequence is synchronized to the EuXFEL 
bunch pattern through the \texttt{SEQUENCER\_CLEAR\_TRIGGER} and 
\texttt{SEQUENCER\_START\_TRIGGER} signals provided by the C$\&$C system, 
respectively. Both trigger signals are sent to Karabo.

In general the pnCCD's sequencer generating the readout sequence and control 
signals for the CAMEX and the CCD can be programmed in a very flexible way, such that the 
CCD's readout and signal processing sequence can be adjusted to and optimized 
for specific experimental requirements. Nevertheless, for most scenarios, changing 
the readout sequence requires a recalibration of the detector for achieving the ideal
performance as shown in Section~\ref{Sec:Detector Performance}.

\section{Data Processing, Correction and Calibration}
\label{Sec:Data Readout and Processing}

The pnCCD online data processing and correction is integrated into Karabo as 
configurable pipelines build of different components, referred to as devices. 
They provide specific signal correction, data processing, data managing and 
data handling functionality. 

Both the digitized analog and pixel clock signals are read out from the 
SIS8300 boards by a C++ software integrated as a device into 
the control system. A second device samples the pixel signal synchronized to
the pixel clock (see Fig.~\ref{Fig:Timing-Diagram}). 
Furthermore, it 
reformats the one-dimensional digitizer data stream into $16$  sub-images with $128 
\times 512$ pixels each (one per CAMEX output channel) and assembles 
them into the full Megapixel image. Since the two digitizer cards are read out 
independently, the two input data streams are synchronized by the timing 
information provided as train IDs in the trigger process. The assembled image 
data is then output as a 2D array with 16~bits resolution per pixel via a 
$10\,\text{Gbit}$ network link to the data acquisition system. Finally, the full 
2D images are stored into HDF5 files, and provided as a data stream to online visualization and 
processing. All software discussed is running on a CPU card hosted in the same 
$\mathrm{\mu}$TCA crate with the SIS8300 boards, communication being done via 
PCI Express.

As offline data processing is not required for rapid feedback and direct 
interfacing to detector hardware, it is conceptually implemented as a stand-alone 
processing pipeline. It is aimed at providing the highest possible data quality, 
i.e. hardware limited calibrated images and spectroscopic data. This pipeline 
is implemented in Python 
and distributed through the pyDetLib library. On the other hand, 
online processing and visualization  is required for rapid feedback to the detector operators 
and users and, therefore, another set of processing pipeline. 

Calibration and correction parameters and their metadata are managed by the 
XFEL calibration data base, stored in HDF5 files and archived on tape. 
Parameters used for data correction are selected by evaluating detector settings 
and conditions under which the data for correction and calibration have been 
acquired. The correction parameter set closest in time to the measured 
scientific data to be corrected and best matching the detector settings is 
finally used. The parameter sets are divided into two different classes, one 
set is provided by the facility and the other set is acquired during the 
user beam time. The latter parameter set might depend and change according to
experiment conditions, while the first parameter set does not 
change significantly on the long term (i.e. on a time scale of weeks to months).

For the pnCCD, the following calibration and correction parameters 
are available for each gain setting through the XFEL calibration data base 
and, if necessary, may be updated before a user's beam time~\cite{Kuster:2014a}:
\begin{itemize}
 \item The gain vectors $g_1, g_{1/4}, g_{1/16}, g_{1/64}, {g_{1/256}}$ and ${g_{1/1024}}$ 
 	for conversion of analog digital units (ADU) to photon numbers,
 \item a vector characterizing the relative variation of the amplification 
	$g_{rel\,i,k}$ 
 	between the 128 different parallel CAMEX readout channels for all 16 ASICs and
 \item the charge transfer inefficiency (CTI) for each sensor column $CTI_{i}$ 
	quantifying the charge loss during charge transfer from the photon interaction 
	point to the readout anode.
\end{itemize}
The following parameters are experiment specific and are determined from data 
acquired during an experiment:
\begin{itemize}
 \item The offset map $O_{i,k}$ providing the per pixel offset,
 \item the noise map providing the per pixel rms noise $N_{i,k}$, and the 
	corresponding bad pixel map $B_{i,k}$.
\end{itemize}

In the following sections we describe the data processing steps applied to pnCCD 
data, as implemented in the online and offline data processing pipelines in more 
detail. The information provided reflects the status as of December 2019, corresponding
to the following software releases: Karabo V 2.4.2, pnCCD calibration pipeline 
pycalibration V 2.0, calPy V 1.9.1-2.4.1, calibration data base cal\_db\_interactive 
V 1.0, the fasterADC V 2.6.1-2.4.0 and cppPnCCDFormatter V 1.1.1-2.4.1 devices.

\subsection{Online Processing}
\label{Sec:Online Processing}

Before scientific data are processed online, a data set for offset and noise 
calculation is typically acquired before each experimental run. The offset and 
noise maps are subsequently calculated per pixel in one single pass using the numerical 
algorithm of \citeasnoun{Welford:1962a}. Welford's algorithm is based on updating the sum 
of squares of differences
\begin{equation}
	M_{2,n} = \sum_{i=i}^n(x_i - \bar{x}_n)^2
\end{equation} 
for estimating the mean of a population of $n$ values. The mean $M_{2,n}$ for the current 
$n$ follows for $n>1$ from 
\begin{equation}
\label{Eq:Mean Value}
	M_{2,n} = M_{2,n-1} + (x_n - \bar{x}_{n-1})(x_n - \bar{x}_{n})
\end{equation} 
and the variance from 
\begin{equation}
	\sigma^2_{n} = \frac{M_{2,n-1}}{n}\,.
\end{equation} 
were $\bar{x}_n$ is the mean calculated from the first $n$ samples.
We calculate the mean signal value of each pixel $i,k$ according to 
equation~\ref{Eq:Mean Value} and assign it to the offset map as value $O_{i,k}$, 
where $i$ and $k$ indicate the column and line numbers of the sensor, $x_n = 
S_{i,k}$ the signal charge measured at pixel position $i, k$ and $n$ the 
number of image frames used for the calculation. In a similar fashion, the noise map 
entry $N_{i,k}$ is calculated from the per pixel variance as $N_{i,k} = 
\sqrt{\sigma^2_n}$. By default, the sample size for mean, variance and standard 
deviation calculation for each pixel is set to $n=500$ images. The resulting offset and noise 
maps are finally injected into the calibration data base. Either this offset map 
is subtracted from the scientific data during online processing or an offset map 
matching the experimental and detector conditions closest is used.

In the latter case, the following detector 
parameters are taken into consideration for selecting the best matching offset 
map: CCD temperature, CCD bias voltage, integration time and the chosen CAMEX gain settings. 
The result of the correction is stored as pixel arrays in 
the corrected data files. During this step, the scientific data are up-converted 
from 16~bit unsigned integer values to 32~bit float values. 

Subsequently, a line-by-line common mode correction reduces baseline variations 
of the analog pixel signal. The number of pixels $N$ in each line taken into 
account for the common mode calculation is configurable to the number of pixels 
per ASIC (i.e. $N=128$ pixels), the dimension of a sensor quadrant ($N=512$ pixels) 
or sensor half ($N=1024$ pixels). The common mode value is then calculated from the 
sorted vector $S_{k,i}$ as
\begin{equation}
	\bar{m}_k = \left\{
	\begin{array}{ll}
	 	S_{k,(N-1)/2} & \mathrm{for\, an \,odd\,number\, of\, pixels} \, N\,\text{in\, line\, k} \\
 		\frac{1}{2} (S_{k,N/2}+S_{k,N/2+1}) & \mathrm{for\, an \,even\,number\, of\, pixels}\, N \,\mathrm{in\,line
 		\,k}
	\end{array}
	\right.
\end{equation}
and subsequently subtracted from the signal in each pixel of one line. 

For the pnCCD it has been found that evaluating the common mode as a thresholded median for 
each quadrant gives a good trade-off between accuracy and stability against event-rich images. 
In that case, only pixels in one line which are carrying a signal below the threshold 
$E_{th}$ are taken into consideration for the calculation of the line's median value.

Further and higher order data treatment like charge sharing, charge transfer 
efficiency corrections and correction of the variations of the relative 
amplification of the 128~CAMEX channels are not taken into account during online 
processing.

The online corrected images are then passed on to the Karabo Graphical User 
Interface (Karabo GUI) for visualization or to the Karabo-Bridge for interfacing 
to user-provided processing using the ZMQ protocol~\cite{Fangohr:2018a}. 

\subsection{Offline Processing}
\label{Sec:Offline Processing}

Offline processing is performed on recorded data files. It is either 
automatically triggered when users migrate data to be archived and made 
available on the Maxwell HPC cluster \cite{Maxwell:2020a}, or explicitly using the Metadata Catalogue 
web interface. Offline processing retrieves necessary calibration and correction 
parameters from the calibration database in a similar way as the online 
processing pipeline. 

By default, the offset correction is always performed using the same offset maps 
as the ones used by the online processing pipeline. Furthermore a line-by-line 
common mode correction using by default the same feature size as for online 
processing is applied. The result of this correction is output into the \texttt{pixel\_cm}
array in the corrected data files. For correcting the data presented in Section~\ref{Sec:Detector Performance}, 
a median value which is calculated from the $N=200$ outermost pixels in each line gives the 
best results with respect to flat field baseline homogeneity. Attention should be paid 
to the fact that the common mode correction  is prone to be biased or can even completely 
fail for very event-rich images. Consequently, the methodology for common mode correction
should be carefully chosen during offline processing.

Noisy pixels, pixels providing no signal (e.g. the area of the central hole) or a 
non-physical signal and events produced by cosmic rays are excluded from the analysis
and marked in the bad pixel map in the calibration data base. Pixels marked as bad are 
excluded from further treatment during the offline processing. 

Split event correction, also called charge sharing correction or charge 
clustering, is always performed. It classifies clustered signals of discrete 
photon events by the number of adjacent pixels the charge is detected in. The signal in 
each pixel is probed against two thresholds prior to event classification: a 
primary event threshold to identify the pixel carrying the majority charge of 
the total signal of a photon event and a secondary (lower) threshold for identifying 
neighboring pixels as having registered a signal above the noise floor. Both 
thresholds are given in units of $\sigma$, where $\sigma$ is the pixels specific 
rms noise given by the noise map. The algorithm then searches for valid charge 
clusters that can originate in a photon interaction and sums all charges above 
noise into an event reconstructed primary pixel carrying the majority 
charge. Clusters that cannot be attributed to a single photon event are 
identified via a special pattern index. Their charge is not summed up. The result of this 
correction is stored into the \texttt{pixel\_classified} array. Additionally, a 
\texttt{patterns\_classified} array is generated, which encodes event multiplicity and 
orientation.

If relative gain correction is activated, the aforementioned
pixel arrays will be stored corrected for the relative difference of the 
amplification of the amplifiers of the 128 channels of each CAMEX.

The Charge Transfer Inefficiency (CTI) of the present generation pnCCDs is 
better than $\approx 5 \times 10^{-5}$, leaving a potential 
signal shift by $2.5\%$ after 512 charge transfer steps. The CAMEX channel to 
channel amplification variation is less than $\pm 5\%$ of the detected signal 
charge. Both corrections are of minor importance for imaging applications and 
are therefore presently neglected during default offline image correction, but can be 
enabled during manual data reprocessing to achieve the best possible spectroscopic 
performance.

Conversion from analog digital units (ADU) to photon numbers is performed 
by means of a gain calibration vector, one for each possible gain configuration 
$g=1, 1/4, 1/16, \ldots 1/1024$. This conversion is performed for the three output
arrays: \texttt{pixels}, \texttt{pixels\_cm} and \texttt{pixels\_classified}. 

For an in depth description of data correction and treatment and its 
implementation of the XFEL pyDetLib, we refer to the corresponding user manuals 
\cite{pyDetLib:2020a}. The data presented in the following sections have been 
treated offline in the same way as described in the previous section.

\section{Detector Performance}
\label{Sec:Detector Performance}


After installation of the detector at the NQS station, we commissioned the pnCCD detector 
and characterized its performance before the first user beamtime, which took place 
shortly thereafter during EuXFEL's user runs in the first and second half of 2019 
(run period~03 and 04 at EuXFEL). We investigated the detector's noise 
behavior and photon response under nominal operating conditions, i.e. at stable 
sensor temperature and optimized detector operating parameters, like bias 
voltage settings, CAMEX configuration and readout sequence of the CCD. The performance 
figures provided in this section refer to the performance as measured 
at the SQS experimental environment.

\begin{figure}
	\centering
  	\includegraphics[width=0.97\columnwidth]{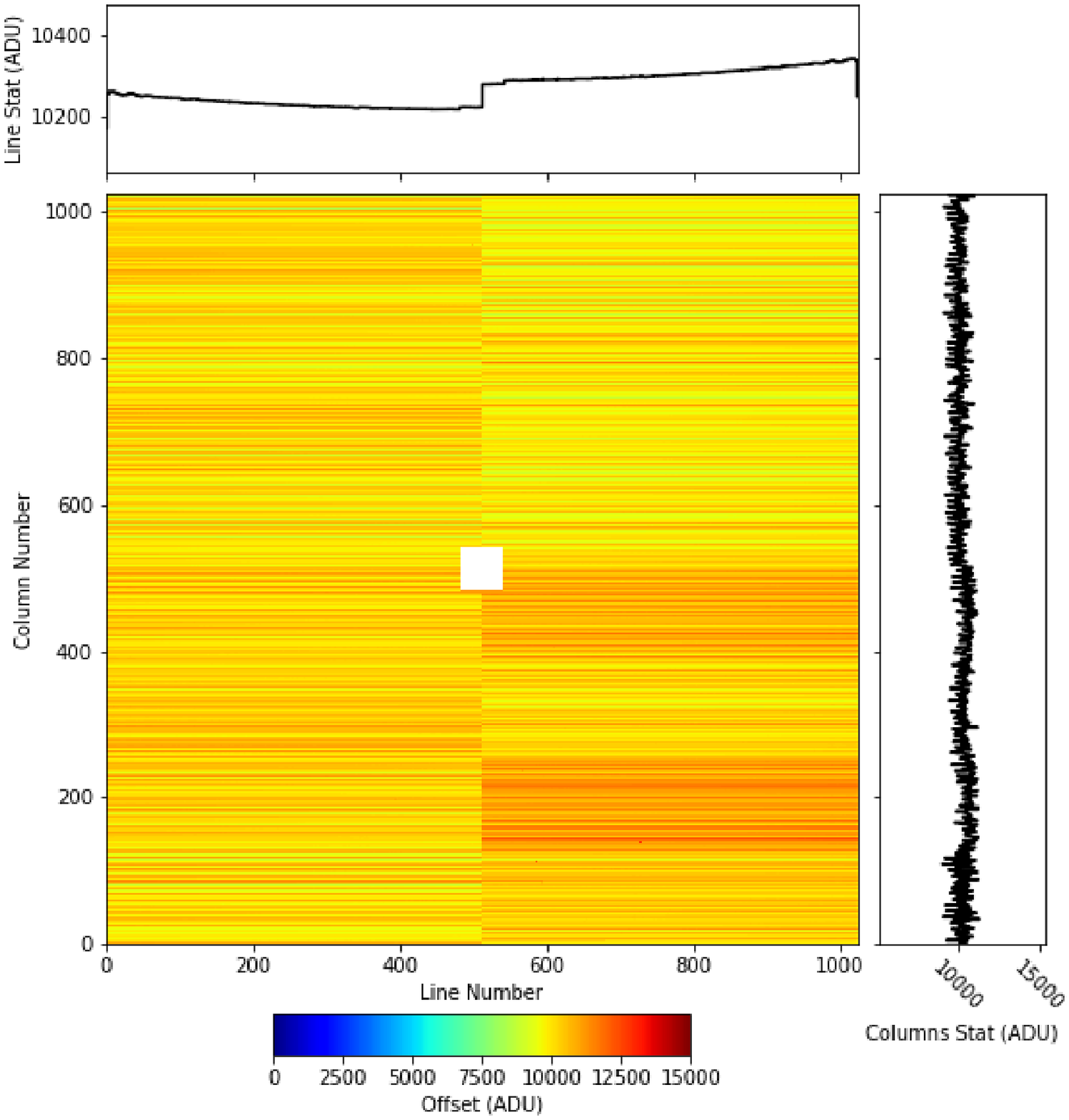}
  	\label{Fig:Sensor-Offset-Gain-16}
	  	\caption{The picture shows the pixel dark signal as measured at the SQS beam line during 
		the first user operation period in $1/16$ gain mode. The dark map is calculated from 400 
		dark frames as described in the text. The orientation of the image is the same as illustrated in 
		Fig.~\ref{Fig:Sensor-Geometry}, i.e. the photon signal is shifted and read out towards the
		CAMEX ASICs located on the left and right side of the sensor. The top and bottom histograms show the 
		offset averaged along the CCD lines (top) and columns (right).}
\end{figure}

Figure~\ref{Fig:Sensor-Offset-Gain-16} shows an exemplary offset map calculated from 
400 individual images acquired with gain $1/16$ during the user 
run period~03. The dark distribution shows the typical regular stripe 
structure along the sensor columns expected from a CCD with column parallel readout. 
This structure can be attributed to channel-to-channel baseline variations of the 
128~CAMEX input channels. 

\begin{figure}
	\centering
 	\includegraphics[width=0.49\columnwidth]{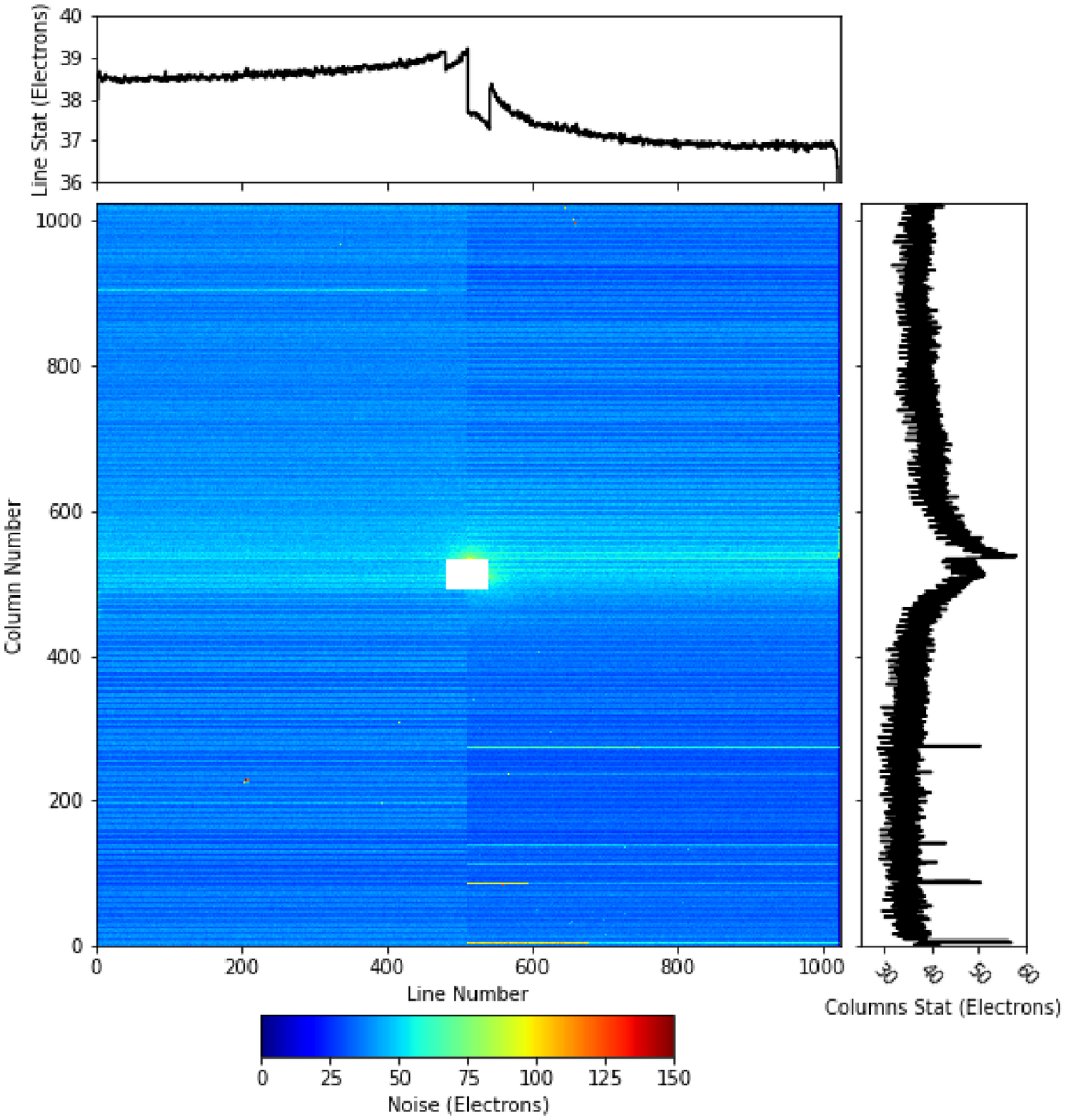}
	\hfill
 	\includegraphics[width=0.49\columnwidth]{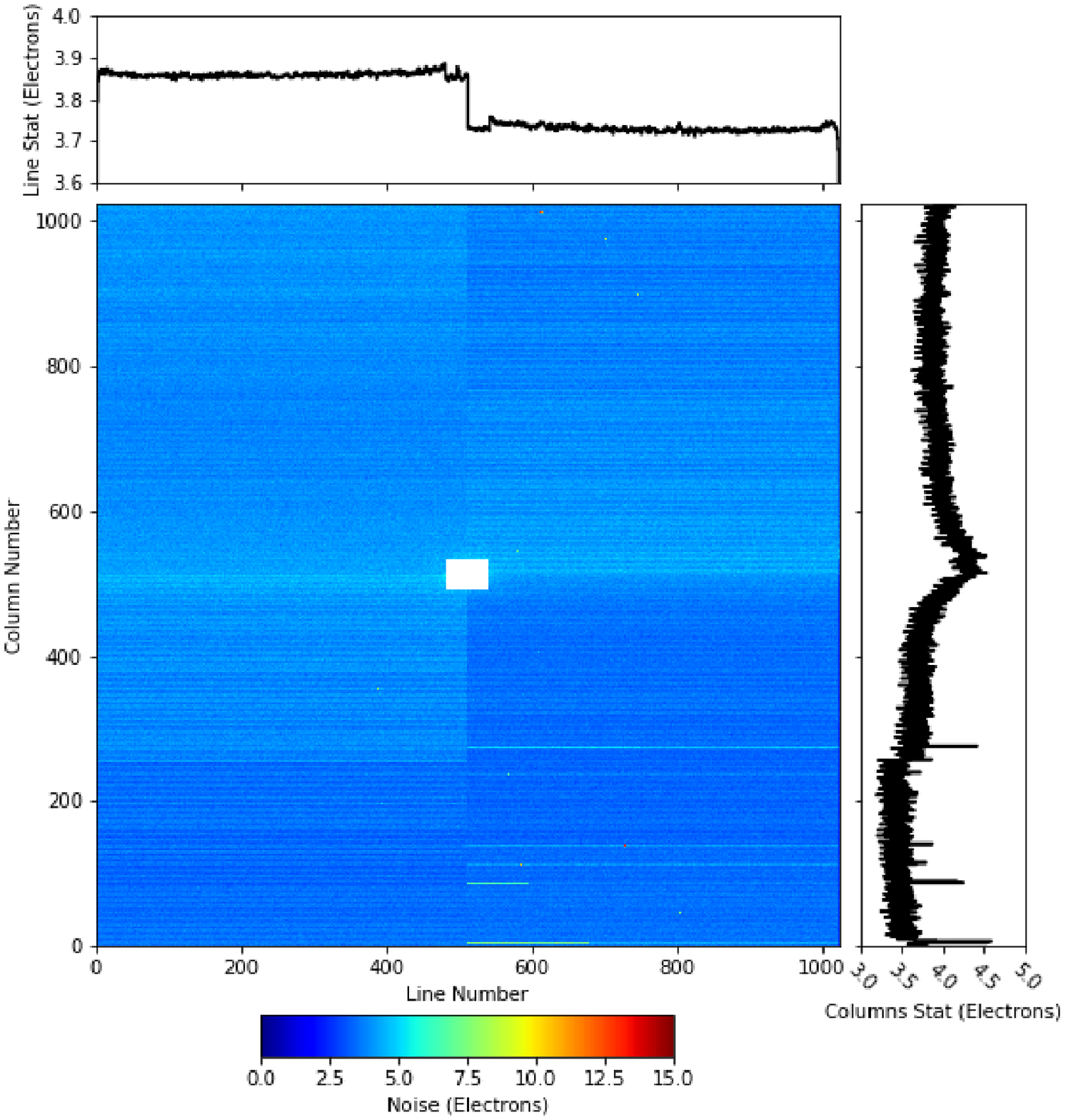}
  	\label{Fig:Sensor-RMS-Noise-Gain-1-16}
	  	\caption{The picture shows the per pixel rms noise as measured at the SQS beam line. 
	  	Left: The per pixel noise measured with gain $1$. Right: The per pixel noise measured with 
	  	gain $1/16$ for comparison.}
\end{figure}

The spatial distribution of the corresponding rms noise is shown in 
Fig.~\ref{Fig:Sensor-RMS-Noise-Gain-1-16} measured with gain $1$ (left image) 
and $1/16$ (right image). 
The features of the dark and noise maps in Fig.~\ref{Fig:Sensor-RMS-Noise-Gain-1-16} can be regarded as exemplary for the operating conditions and parameter settings used with the pnCCD detector during the EuXFEL's user beam time.

As mean dark signal calculated across the whole chip we find $10968.7\,\text{ADU}$. 
The corresponding per pixel noise of $126.15\,\text{ADU} = 37.9\,\text{eV}$, translates into an 
equivalent noise charge (ENC) of $10.2\,\text{e}^{-}$, when taking the ADU to photon energy 
conversion factor of $0.30\,\text{eV}/\text{ADU}$ (gain $1$) into consideration. 
The EuXFEL pnCCD detector is thus capable of discriminating single photons with an energy 
of $250\,\text{eV}$ and $500\,\text{eV}$ from the noise floor with a significance 
of $6.6\,\sigma$ and $13.2\,\sigma$, respectively.

\begin{figure}
	\centering
  	\includegraphics[width=0.97\columnwidth]{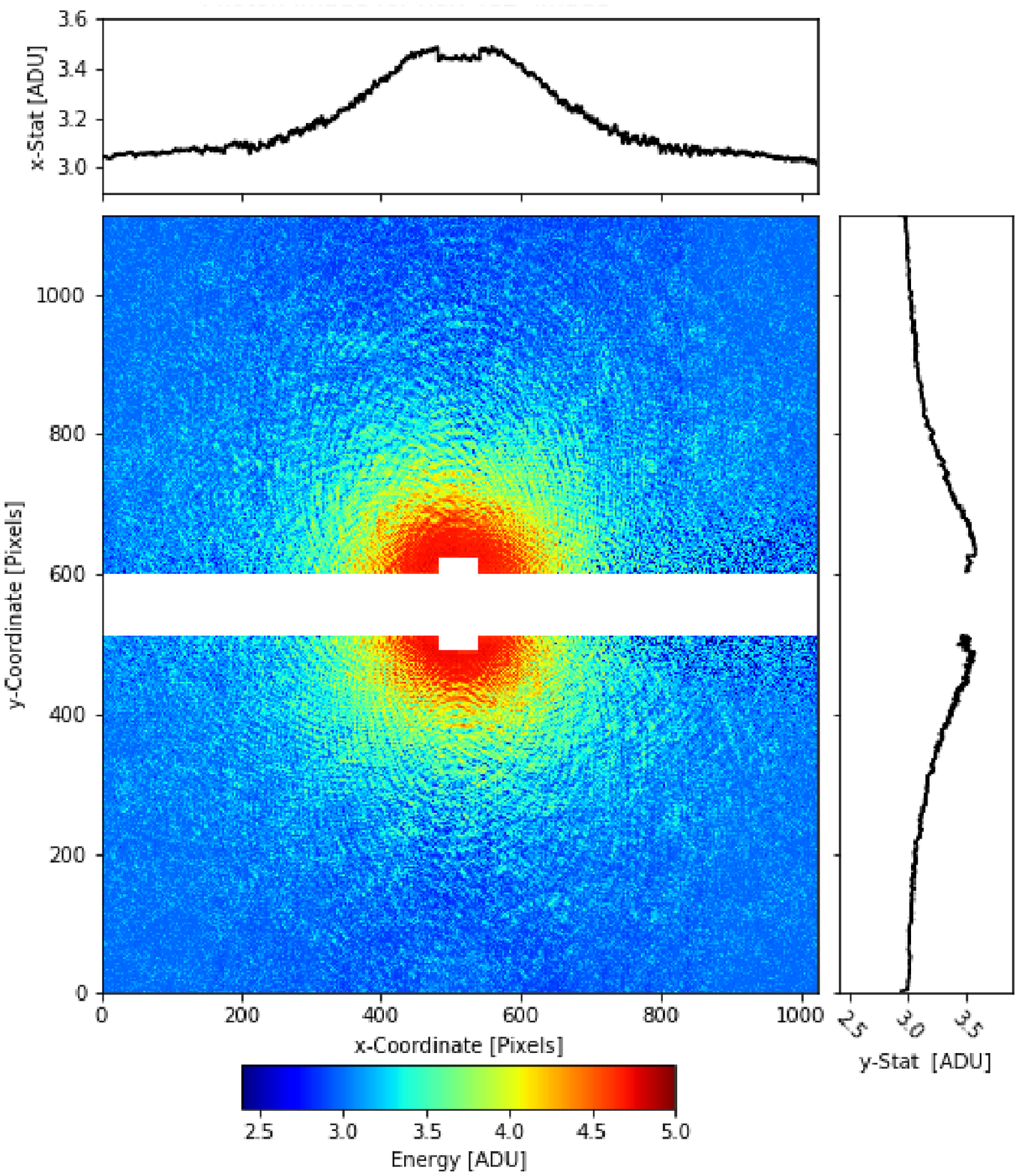}
  	\vspace{0.7cm}
  	
  	\includegraphics[width=0.49\columnwidth]{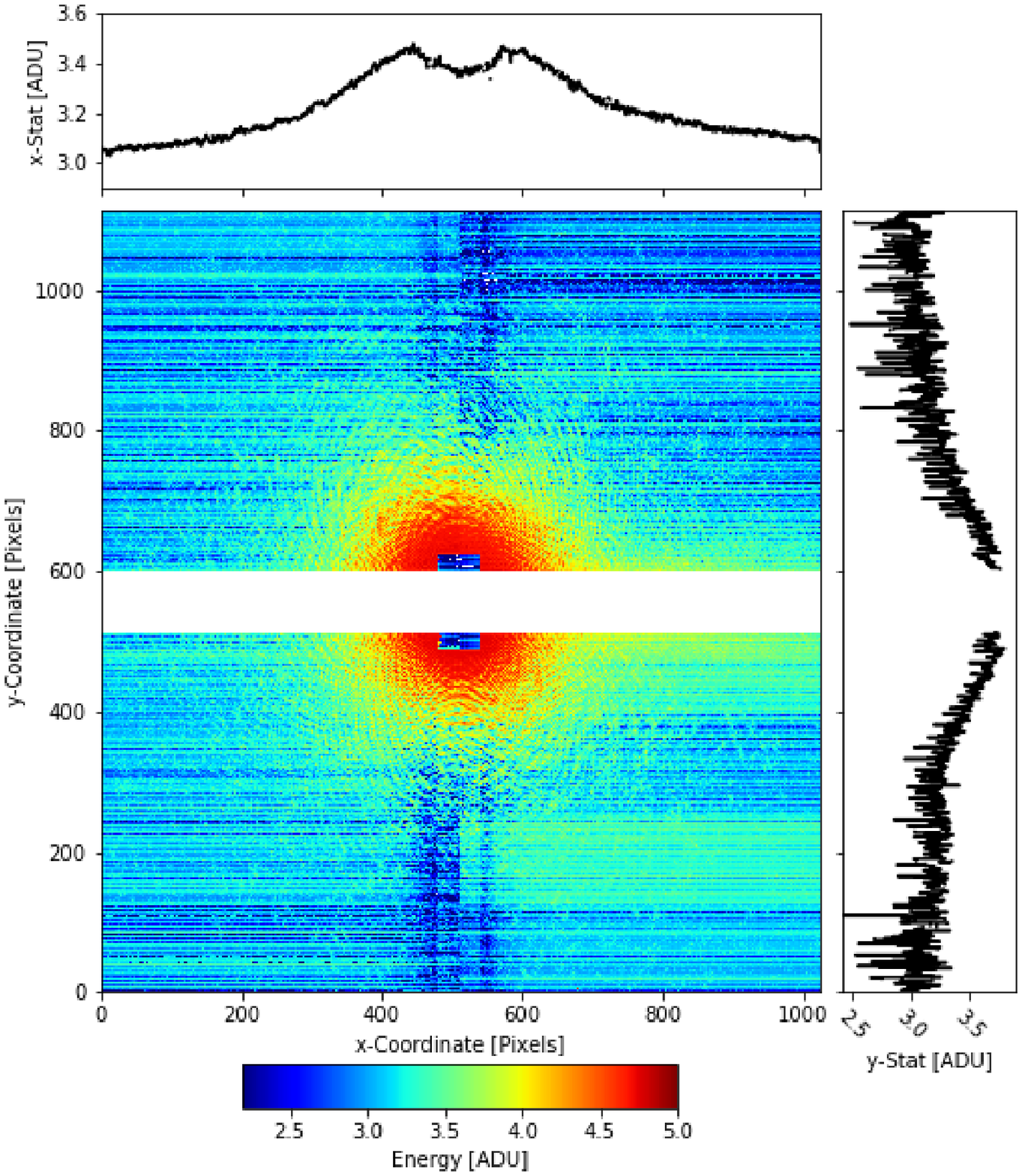}
  	\hfill
  	\includegraphics[width=0.49\columnwidth]{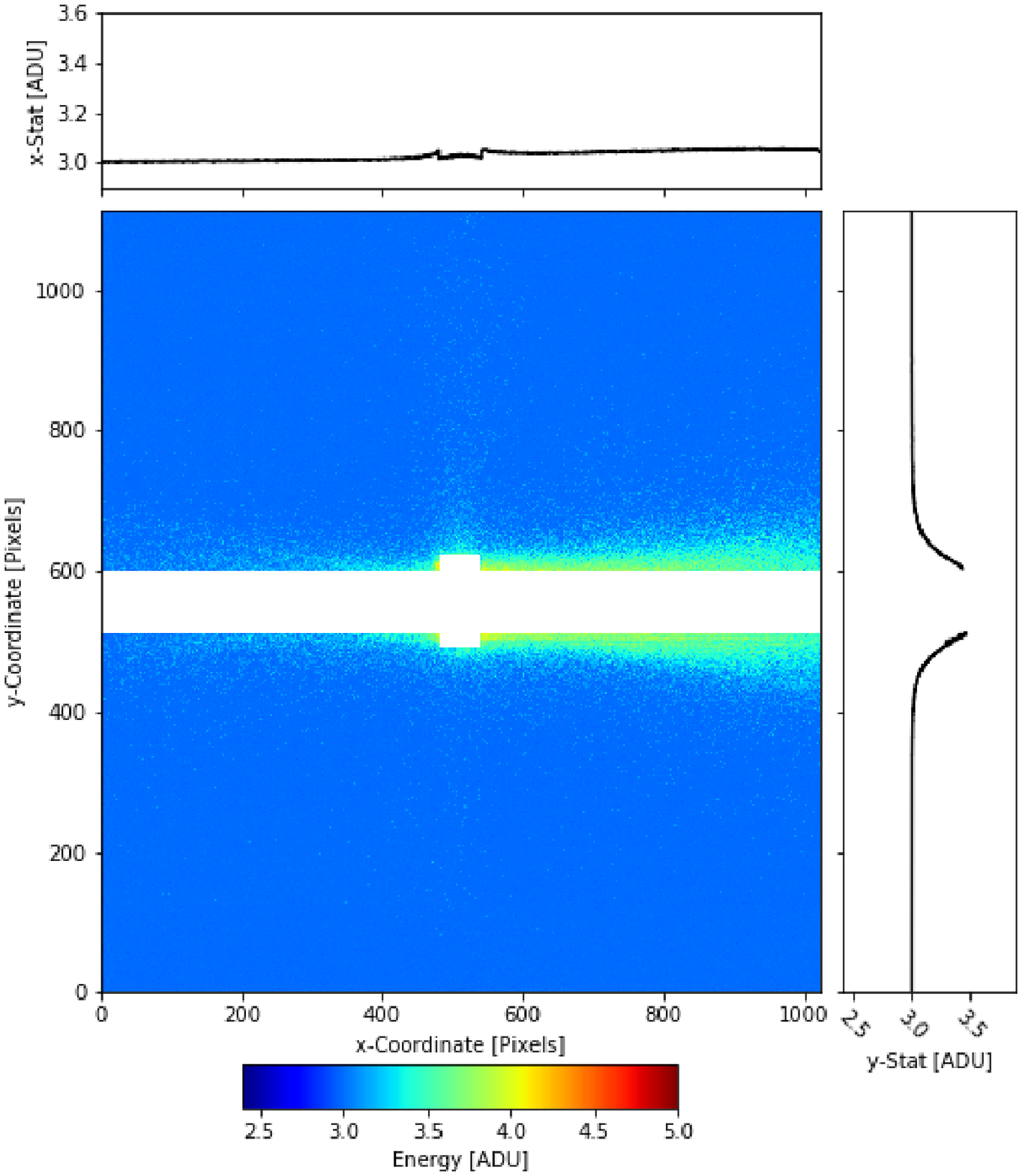}
  	\label{Fig:Sensor-Difffraction-Pattern2}
  	
	\caption{Top: Fully corrected single-shot coherent X-ray diffraction image of a 
		superfluid helium droplet doped with acetonitrile molecules. The diffraction 
		image was obtained from a single FEL pulse at a photon energy of $1\,\text{keV}$ and 
		was recorded in June 2019 with the newly-commissioned pnCCD sensor at the NQS chamber. 
		During data acquisition, the detector's gain was set to $1/16$ and the top and bottom 
		CCD modules were separated by $6.8\,\text{mm}$ vertically. Bottom left: 
		The uncorrected raw image in units of ADU as provided by the output of the ADC is shown. Bottom 
		right: Experiment background image obtained from a single-shot, acquired directly before the signal 
		image.}
\end{figure}

An example of the detector's imaging performance is demonstrated by 
Fig.~\ref{Fig:Sensor-Difffraction-Pattern2}. This diffraction pattern was 
acquired during the first user beam time using the pnCCD installed at the NQS 
chamber in June 2019~\cite{Tanyag:2020a}. The diffraction image was obtained from a micron-sized 
superfluid helium droplets containing millions of acetonitrile molecules. The droplets 
were hit by one single FEL pulse focused to a diameter of about $1.5\,\mu\text{m}$, 
with a photon energy of $1\,\text{keV}$, 

The detector was configured to gain $1/16$, in order to avoid saturation of the 
pixels close to the central imaging area. The area of the central hole is 
excluded from the analysis. Further data treatment includes subtracting the 
experimental background captured by the imaging frame before the helium 
diffraction pattern (bottom left picture of Fig.~\ref{Fig:Sensor-Difffraction-Pattern2}).

The influence of the data corrections applied during offline analysis is 
illustrated by comparing the bottom left picture of 
Fig.~\ref{Fig:Sensor-Difffraction-Pattern2} with the fully corrected diffraction 
pattern on the top. The bottom left picture shows the same diffraction pattern 
as provided by the front end electronics at the output of the ADC. Apart from 
normalization of the mean baseline to the baseline of the corrected image, no further 
data corrections were applied to this image. The uncorrected image shows all 
features of the detector's dark signal, common mode effects, experiment 
background and the diffraction signal on top. The horizontal stripes result from 
variations in the baseline and gain for each CAMEX readout channel. Furthermore, 
a non-zero signal is apparent from the area of the central hole. This is the 
consequence of the way the pnCCD sensor is read out. Since the readout sequence 
always considers $512$ pixels to be processed by the CAMEX, the CAMEX provides
an output signal to the ADC even for the area corresponding to the central hole.


\section{Conclusions and Outlook}
 
The EuXFEL 1-Megapixel pnCCD detector enables experimental techniques like 
coherent diffraction imaging at the SQS instrument with single photon 
sensitivity on a $6.6\,\sigma$ and $13.2\,\sigma$ significance level at the lowest 
photon energies provided by the FEL, i.e. $250\,\text{eV}$ in the future and 
presently $500\,\text{eV}$. 

The detector is installed at the EuXFEL's SQS instrument and has successfully 
been integrated into the NQS endstation, tested and subsequently commissioned in 
May 2019. The fully equipped NQS endstation is available for the European XFEL's 
user community since June 2019 for investigations of the interaction of intense 
FEL pulses with clusters, nano-particles and small bio-molecules exploiting 
coherent diffraction imaging techniques. Since then, the pnCCD detector and the 
NQS endstation have already been successfully used by the scientific community for 
several experiments during user runs in the first and second half of 2019 at EuXFEL.

The detector system is integrated into the EuXFEL's facility-wide control 
system Karabo, enabling the EuXFEL standardized control, online/offline data 
processing, on the fly data correction and data calibration. During the 
commissioning phase and first user experiments, stable operation with a noise 
level below $10\,{e^-}$ ENC was demonstrated (see Tab.~\ref{Table:pnCCD-Detector-Parameters}). 

Further improvements of the performance of the detector are expected after the 
present sensor module has been replaced by a further optimized one. This major 
upgrade is planned for the second half of 2020. The new sensor module will 
provide an improved sensitivity for low energy photons, enabled by the 
implementation of a thinner aluminum entrance window with a thickness of 
$60\,\text{nm}$ in comparison to $150\,\text{nm}$ of the present one. This 
results in a factor of $1.5$ higher quantum efficiency at $0.25\,\text{keV}$. An 
extensive multi-energy calibration campaign of the pnCCD will follow this
upgrade, with the main goals of enabling best possible sensitivity and the 
spectroscopic capabilities of the pnCCD with Fano-Noise-limited energy 
resolution.

Further information about the SQS instrument, the NQS station, data 
processing/correction, and the pnCCD detector can be found on the EuXFEL's 
webpages \cite{SQS:2020a}.


\appendix



\ack{Acknowledgement}
We acknowledge European XFEL in Schenefeld, Germany, for provision of X-ray free-electron 
laser beamtime at the SQS instrument and would like to thank the staff for their assistance.
The work presented in this publication was funded by EuXFEL. D.R., R.T., and A.U. 
acknowledge funding from the Bundesministerium für Bildung und Forschung via grant 
No. 05K16KT3, within the BMBF Forschungsschwerpunkt Freie-Elektronen-Laser FSP-302 
and from the Leibniz-Gemeinschaft via grant No. SAW/2017/MBI4. We would like to thank 
specifically the following EuXFEL groups for their fruitful collaboration, vital 
contribution to this work and their continuous effort in supporting this project: Control 
devices were developed by the Control and Analysis Software (CAS) group led by 
Sandor Brockhauser, the motor control hardware, timing system and PLC systems 
were contributed by the Electronic \& Electrical Engineering (EEE) group led by 
Patrick Gessler, data acquisition and storage is provided by the Information 
Technology and Data Management (ITDM) group led by Krzystof Wrona.




\bibliography{sqs-pnccd}
\bibliographystyle{iucr}



\begin{table}
  \caption{Detector performance parameters as achieved at the SQS experiment in comparison to 
  	detector parameters.}
  \label{Table:pnCCD-Detector-Parameters}
  \begin{tabular}{lrr}      
 	Detector Parameter                   & Achieved at SQS                          \\ \hline
 	Energy range                         & $0.5$ -- $3\,\text{keV}$                 \\
 	Pixel size                           & $75\,\mu\text{m} \times 75\,\mu\text{m}$ \\
 	Position resolution                  & $< 10\,\mu\text{m}$                      \\
 	Sensor thickness                     & $450\,\mu\text{m}$                       \\
	Charge handling capacity             & $\approx 5 \times 10^5\,e^-/\text{pixel}$\\
 	Sensitive area (both sensor halves)  & $7.7 \times 7.7\,\text{cm}^2$            \\
 	Dynamic range (low gain)             & $6000\,\text{ph} @1\,\text{keV}$         \\
 	RMS noise                            & $10.2\,e^-$ (gain 1/16, low gain)        \\
									 	& $37.9\,e^-$ (gain 1, high gain)          \\
 	Single photon sensitivity            & $6.6\sigma$ at $\ge 250\,\text{eV}$      \\
 	                                     & $13.2\sigma$ at $500\,\text{eV}$         \\
 	Max. frame rate                      & up to $100\,\text{Hz}$                   \\
  \end{tabular}
\end{table}


\end{document}